\documentclass[conference]{IEEEtran}
\IEEEoverridecommandlockouts
\usepackage{cite}
\usepackage{amsmath,amssymb,amsfonts}
\usepackage{algorithmic}
\usepackage{graphicx}
\usepackage{textcomp}
\usepackage{xcolor}
\usepackage{float}
\usepackage{booktabs}
\def\BibTeX{{\rm B\kern-.05em{\sc i\kern-.025em b}\kern-.08em
    T\kern-.1667em\lower.7ex\hbox{E}\kern-.125emX}}
\begin{document}

\title{E2CAR: An Efficient 2D-CNN Framework for Real-Time EEG Artifact Removal on Edge Devices\\
\thanks{$^{\star}$Equal contribution.}
\thanks{$^{\dagger}$Corresponding Author}
}


\author{
    \IEEEauthorblockN{
        Haoliang Liu\IEEEauthorrefmark{1}$^{\star}$, 
        Chengkun Cai\IEEEauthorrefmark{2}$^{\star}$, 
        Xu Zhao\IEEEauthorrefmark{2}$^{\star}$, 
        Lei Li\IEEEauthorrefmark{3}\IEEEauthorrefmark{4}$^{\dagger}$
    }
    \IEEEauthorblockA{
        \IEEEauthorrefmark{1}University of Manchester 
        \IEEEauthorrefmark{2}University of Edinburgh \\
        \IEEEauthorrefmark{3}University of Washington 
        \IEEEauthorrefmark{4}University of Copenhagen \\
        \small
        haoliang.liu@postgrad.manchester.ac.uk, 
        \{C.Cai-13, X.Zhao-95\}@sms.ed.ac.uk, 
        lilei@di.ku.dk
    }
}

\maketitle

\begin{abstract}
Electroencephalography (EEG) signals are frequently contaminated by artifacts, affecting the accuracy of subsequent analysis. Traditional artifact removal methods are often computationally expensive and inefficient for real-time applications in edge devices. This paper presents a method to reduce the computational cost of most existing convolutional neural networks (CNN) by replacing one-dimensional (1-D) CNNs with two-dimensional (2-D) CNNs and deploys them on Edge Tensor Processing Unit (TPU), which is an open-resource hardware accelerator widely used in edge devices for low-latency, low-power operation. A new Efficient 2D-CNN Artifact Removal (E2CAR) framework is also represented using the method above, and it achieves a 90\% reduction in inference time on the TPU and decreases power consumption by 18.98\%, while maintaining comparable artifact removal performance to existing methods. This approach facilitates efficient EEG signal processing on edge devices.
\end{abstract}


\section{Introduction}

EEG signals are vital for understanding brain activities, but they are often contaminated by various artifacts such as ocular, muscular, and environmental noise \cite{chaddad2023electroencephalography,jiang2019removal}. The removal of these artifacts is critical for accurate EEG analysis. Traditional methods, such as Independent Component Analysis (ICA) and Canonical Correlation Analysis (CCA), have limitations regarding computational complexity, real-time applicability, and the need for domain expertise\cite{djuwari2006limitations,naik2006limitations,mumtaz2021review}. With the advent of deep learning, neural networks have shown significant promise in enhancing artifact removal from EEG signals by leveraging their ability to learn complex, non-linear patterns\cite{mashhadi2020deep,stalin2021machine,yang2018automatic}.

\begin{figure}[ht]
    \centering
    \includegraphics[width=0.48\textwidth]{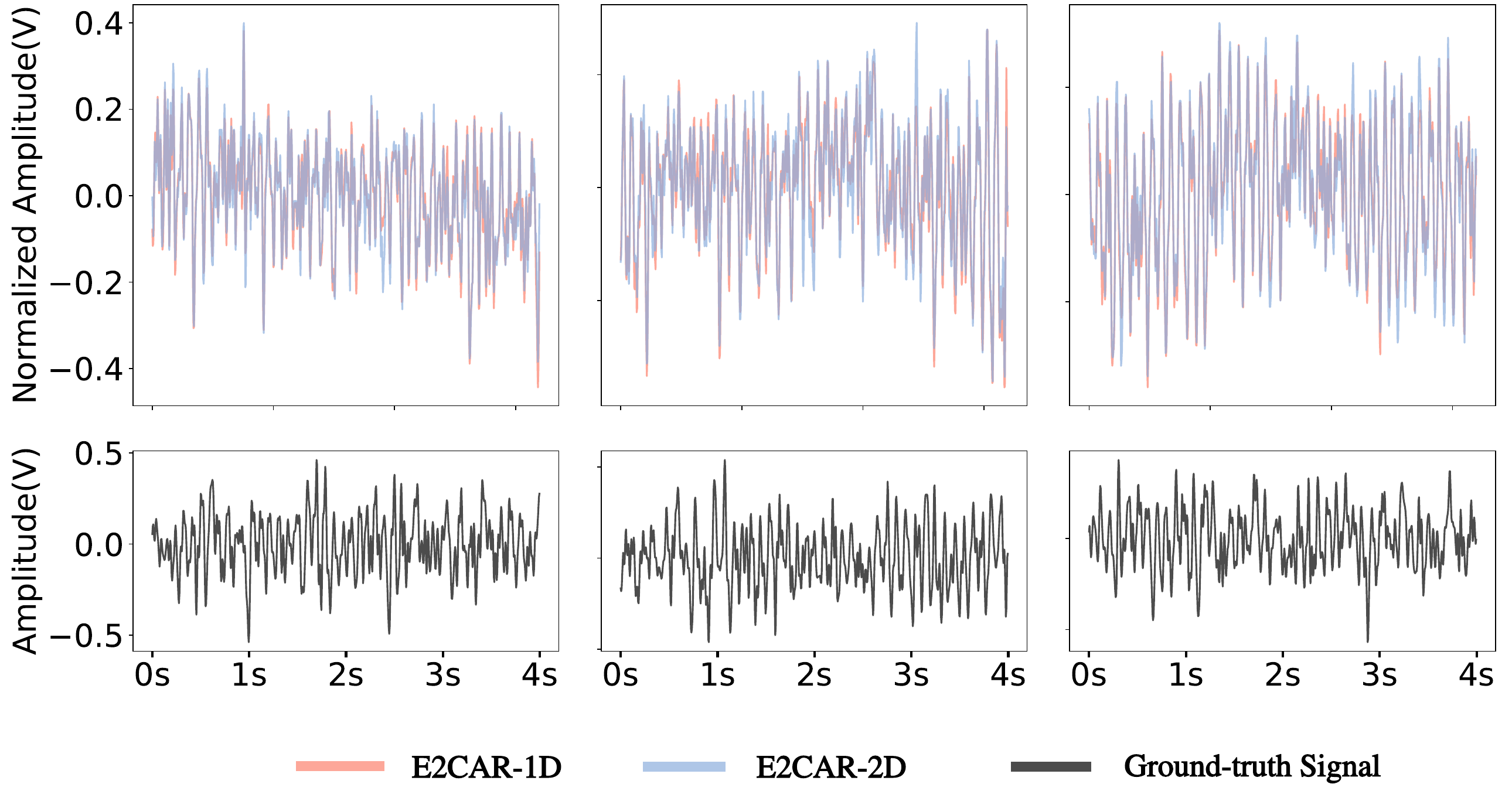}
    \caption{E2CAR output performance on Coral Dev board mini}
    \label{fig:output}
\end{figure}

Despite these advancements, deploying deep learning models on edge devices for real-time applications poses substantial challenges, primarily due to computational resource constraints and power limitations\cite{zhang2020deep,chen2020deep}. And most existing models share a common feature: adopting 1-D CNNs, which is intuitively correct but not suitable for edge devices. To address these issues, this work introduces a novel approach by converting 1-D CNN to 2-D CNN, since most of the work relating with EEG artifact removal is using 1-D CNN, combining with hardware accelerator to reduce computational cost.  This work also modifies a 1-D deep autoencoder\cite{xing2024deep} (DAE) model by adding a residual Convolutional Neural Network \cite{he2016deep} (Reset) module to improve its performance. The new model is applied to Google's Coral Dev Board mini, utilizing its edge TPU as a hardware accelerator to improve computational efficiency and reduce power consumption. One example of the artifact removal performance is shown in Figure\ref{fig:output}.

The contributions of this study are threefold: (1) proposing an optimization method to improve most of the EEG artifact removal models for applying these models on edge devices, (2) proposing a new model structure and using the optimization method, and (3) evaluating new model's performance in terms of artifact removal accuracy, inference time, and power consumption compared to existing models.

\section{Related work}

\subsection{EEG Artifact Removal}
Traditional methods for removing artifacts from EEG signals include Independent Component Analysis (ICA) \cite{vigario1997extraction}, Canonical Correlation Analysis (CCA) \cite{zhuang2020technical}, and Blind Source Separation (BSS) \cite{fitzgibbon2007removal}. These approaches aim to decompose mixed EEG recordings into latent source components under specific statistical assumptions, such as source independence or low inter-channel correlation, and have been widely adopted for mitigating ocular and muscle artifacts. However, such classical techniques typically require manual intervention (e.g., component selection) and rely on assumptions that may not hold consistently across subjects, recording conditions, and sensor configurations, which can lead to unstable performance in practice.

To overcome these limitations, recent studies have incorporated machine learning and deep learning techniques to automate EEG artifact removal, including convolutional neural networks (CNNs) \cite{zhang2021eegdenoisenet}, residual networks (ResNet) \cite{he2016deep}, and denoising autoencoders (DAE) \cite{xing2024deep}. By learning representations directly from data, these methods reduce reliance on handcrafted features and explicit source separation assumptions, enabling improved robustness to complex and non-stationary artifacts. This shift from assumption-driven signal decomposition toward data-driven modeling reflects a broader methodological transition observed in modern intelligent systems, where inductive reasoning increasingly complements or replaces deductive formulations \cite{cai2025role}.

EEG artifact removal further requires effective modeling of long-range temporal dependencies under noisy conditions. Recent work on time-series modeling has shown that model capacity and scaling behavior play an important role in capturing structured temporal dynamics, especially when signals are contaminated by noise \cite{shi2024scaling}. In addition, motion-related artifacts in EEG often exhibit structured and repetitive temporal patterns induced by body movement and muscle activity. Advances in human motion modeling similarly emphasize the importance of structured temporal representations for handling complex motion-induced signal variations \cite{li2025human}. 

Beyond performance considerations, interpretability and transparency have gained increasing attention in learning-based models, particularly for safety-critical applications involving complex temporal signals \cite{lan-etal-2025-attention}. Moreover, optimizing model behavior under noisy feedback and multiple evaluation criteria has been explored in other domains using principled optimization frameworks, highlighting complementary perspectives for robust model tuning in artifact-prone settings \cite{cai2025bayesian}. These trends motivate the design of lightweight and structured architectures that balance denoising effectiveness, robustness, and deployability.

\begin{figure*}[ht]
    \centering
    \includegraphics[width=\textwidth]{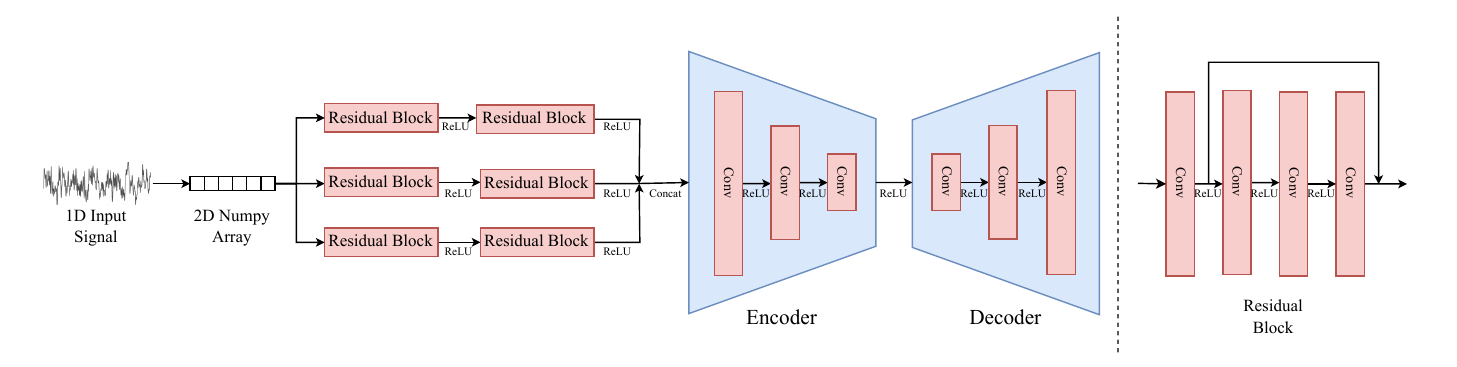}
    \caption{Overview of the proposed E2CAR framework, where the 2D-CNN autoencoder with residual connections processes EEG signals for efficient feature extraction and artifact suppression while preserving essential signal characteristics.}
    \label{fig:architecture}
\end{figure*}

\subsection{Embedded Application for EEG}
Deploying deep learning models for EEG processing on embedded or edge devices has attracted growing interest due to practical requirements such as portability, privacy preservation, and real-time responsiveness. Prior studies have explored efficient neural architectures and deployment strategies for resource-constrained platforms, including compact CNN and ResNet variants as well as autoencoder-based models \cite{canziani2017evaluation,yu2022edge}. Hardware accelerators such as the Edge TPU further enable low-latency and energy-efficient inference through quantized execution, making them suitable for continuous EEG monitoring applications \cite{garcia2023analysing}. Benchmarking efforts also indicate that quantization-aware design and operator efficiency are crucial for maintaining favorable accuracy--efficiency trade-offs on edge devices \cite{baller2021deepedgebench}.

From a modeling perspective, many EEG artifacts—particularly those induced by motion and muscle activity—exhibit repetitive and structured temporal patterns. Related research in other domains has investigated how to capture repetitive temporal dynamics efficiently and robustly, emphasizing compact representations and strong generalization under limited computational budgets \cite{yao2025countllm,gu2025mocount}. Although these works focus on different sensing modalities, the underlying challenge of learning structured temporal cues under noise and resource constraints is closely aligned with embedded EEG artifact removal. Building on these insights, this work optimizes a 2D-CNN autoencoder architecture specifically for TPU deployment, enabling low-latency and energy-efficient EEG artifact removal in constrained environments.


\section{Methodology}
\subsection{Data Preparation and Pre-processing}
In this study, a standardized data preprocessing pipeline was applied to address different types of EEG artifacts, including Electrooculography (EOG), motion artifacts, and Electromyography (EMG), ensuring that the model could be trained and inferred under consistent input conditions. For EOG data, the dataset in \cite{klados2016semi} was used. It contains 54 pairs of clean and corrupted EEG recordings from 27 participants. For motion artifact data, the dataset in \cite{sweeney2012methodology}was used, which contains 23 pairs of EEG recordings. As to EMG contaminated data, the dataset in \cite{zhang2021eegdenoisenet} was used. This dataset contains 5598 EMG artifact segments. This work uses the same data processing method mentioned in paper\cite{xing2024deep}. First, all EEG signals underwent down sampling and detrending to unify the sampling frequency across datasets to 200 Hz and remove low-frequency drifts. Then, for the clean versions of the EEG signals, which is also used for EEG clean singal reconstruction task, a 1-50 Hz band-pass filter was applied to eliminate high-frequency noise and low-frequency interference. For the artifact-contaminated EEG data, no filtering was applied, preserving the original characteristics for the model to learn from the raw artifact data. To avoid edge artifacts introduced by filtering, 2 to 5 seconds of data were removed at the beginning and end of each recording. Next, all EEG data were segmented into 4-second period with 50 percent overlap between adjacent segments, which increased the data volume and enhanced the model's ability to capture temporal features. Finally, min-max normalization was applied to each segment individually, scaling the signal amplitude to a range between 0 and 1 to eliminate the impact of amplitude differences during model training. This normalization process was also applied during real-time inference, ensuring efficient performance on mobile devices such as smartphones.

\subsection{Model Structure}
DAE model proposed by paper \cite{xing2024deep} fail to converge in the training progress. To address the vanishing gradient problem and improve the model's ability to learn diverse features from the signal, six residual blocks, arranged in a 3×2 configuration, are integrated before the encoder, as illustrated in Figure \ref{fig:architecture}. The residual blocks are consisted of 1×3, 1×5, 1×7 converlution kernels \cite{he2016deep}. These blocks can significantly enhance the model's feature extraction capabilities while effectively mitigating the vanishing gradient issue, which is often encountered in deep networks. The Residual blocks introduce shortcut connections, allowing gradients to bypass multiple layers and flow more directly, thus improving training efficiency and convergence. The different size of kernels are used to extract features from different artifacts. EOG artifacts are slow-varying artifacts, usually caused by blinking or eye movement. These artifacts have low frequencies and long time spans. So using large converlution kernel(1×7) can extract the slow-varying patterns of the artifacts, distinguishing them from fast-changing signals. On the contrary, EMG artifacts are caused by facial or head muscle activity and typically manifest as high-frequency oscillations over a short period, with relatively low amplitude but high frequency, so a 1×3 kernel is perfect for this task. When it comes to motion artifact, which have a broad spectrum of temporal characteristics, all the small, medium(1×5), and large kernels work. This modification enables the model to capture a wide variety of features caused by different factors during EEG signal collection.

\subsection{Model Optimization}
Moreover, to solve the issue of inference time exceeding the duration of signal collection, the model architecture has been expanded from processing 1D signals directly to using 2D numpy arrays, as illustrated in the figure\ref{fig:architecture}. In this work, EEG data collected by sensors consists of 800 data points, which are directly fed into a 1D CNN. This approach is currently adopted by all existing studies. Researchers assume by default that these data points are arranged in chronological order, thereby implicitly hiding the time dimension. In contrast, the method proposed in this paper explicitly represents the time dimension by transforming the temporal signals into an image, with the horizontal axis representing time and the vertical axis representing signal values. This image with a size of 1×800 is then fed into a 2D CNN.

Modern embedded CPUs, such as the ARM Cortex-M7 CPU used in devices like the Coral Dev Board, are optimized for 2D operations. These CPUs integrate DSP instruction sets that optimize dot product and accumulation operations for 2D matrices, whereas 1D operations often require additional memory access, reducing efficiency. Additionally, memory access efficiency and parallel computation are critical for speeding up inference. TPUs, which are optimized for matrix multiplication tasks, handle 2D convolution more efficiently, allowing for better parallelization and maximizing hardware usage. As a result, adopting 2D convolution significantly improves inference speed, especially on TPUs, compared to 1D convolution, which requires additional steps to simulate matrix operations.


\subsection{Evaluation Metrics}

\subsubsection{Relative Root-Mean-Square Error}
RRMSE evaluates the difference between reconstructed and ground-truth EEG in both the time and frequency domains:

\begin{equation}
RRMSE_t = \frac{RMS(y - \tilde{x})}{RMS(\tilde{x})}
\end{equation}

\begin{equation}
RRMSE_f = \frac{RMS(PSD(y) - PSD(\tilde{x}))}{RMS(PSD(\tilde{x}))}
\end{equation}

where \( y \) is the reconstructed EEG, \( \tilde{x} \) is the ground-truth EEG, and PSD represents the Power Spectral Density.

\subsubsection{Correlation Coefficient}

The formula for Correlation Coefficient (CC) quantifies the linear relationship between two variables, the ground-truth EEG signal $y$ and the reconstructed EEG signal $\tilde{x}$. It is expressed as:

\begin{equation}
CC = \frac{Cov(y, \tilde{x})}{\sqrt{Var(y) \cdot Var(\tilde{x})}}
\end{equation}
where
\begin{itemize}
    \item $Cov(y, \tilde{x})$ is the covariance between the ground-truth signal $y$ and the reconstructed signal $\tilde{x}$, which measures how much the two signals vary together.
    \item $Var(y)$ and $Var(\tilde{x})$ are the variances of the ground-truth and reconstructed signals, respectively, representing the amount of variation within each signal.
\end{itemize}

A higher CC value, closer to 1, indicates a strong linear relationship between the two signals, suggesting the reconstructed signal closely matches the original. This implies effective artifact removal with minimal distortion to the EEG signal. Conversely, a lower CC value suggests weaker correspondence, indicating potential signal loss or distortion.

\subsubsection{Computational Time and Power Consumption}
Inference time can be obtained using time function in Python. We use the time value when the model gives an output minus the time value when the model starts inference. In this way we can calculate the time different model takes to run inference on Coral Dev board mini. The power consumption is measured using a power meter. The unit of power consumption is mAh/h. To obtain the power consumed by the model during operation, subtract the charge consumed at the end of the model's operation from the charge consumed during operation, and divide by the time, which can be represented as:

\[
P = \frac{Q_{end} - Q_{start}}{t}
\]

where
\begin{itemize}
    \item $P$ is the power in mAh/h,
    \item $Q_{end}$ is the charge at the end of the model's operation,
    \item $Q_{start}$ is the charge at the start of the model's operation,
    \item $t$ is the time duration of the operation (in hours).
\end{itemize}

\begin{table*}[ht]
\centering
\label{table:clean_rrmse}
\caption{Clean Signal Reconstruction - Comparison of RRMSE values and CC}
\begin{tabular}{@{}lccccccccccc@{}}
\toprule 
\textbf{Metric} & \textbf{CNN} &
\textbf{CNN-2D} & \textbf{ResNet} & \textbf{ResNet-2D} & \textbf{DAE} & \textbf{DAE-2D} & \textbf{E2CAR} & \textbf{E2CAR-2D} & \textbf{E2CAR-TPU} \\ 
\midrule
RRMSE-Time & 0.39 ± 0.05 & 0.56 ± 0.05 & 0.37 ± 0.05 & 0.31 ± 0.03 & 0.29 ± 0.07 & 0.28 ± 0.07 & 0.23 ± 0.07 & 0.30 ± 0.06 & 0.30 ± 0.06 \\
RRMSE-Freq & 0.46 ± 0.10 & 0.68 ± 0.06 & 0.46 ± 0.14 & 0.47 ± 0.09 & 0.35 ± 0.12 & 0.35 ± 0.12 & 0.17 ± 0.11 & 0.36 ± 0.10 & 0.36 ± 0.10 \\
CC & 0.93 ± 0.02 & 0.83 ± 0.04 & 0.95 ± 0.01 & 0.97 ± 0.01 & 0.96 ± 0.02 & 0.96 ± 0.02 & 0.97 ± 0.02 & 0.96 ± 0.02 & 0.95 ± 0.02 \\ 
\bottomrule
\end{tabular}
\end{table*}

\begin{table*}[ht]
\centering
\label{table:eog_rrmse}
\caption{EOG Artifact Removal - Comparison of RRMSE values and CC}
\begin{tabular}{@{}lccccccccccc@{}}
\toprule
\textbf{Metric} & \textbf{CNN} & \textbf{CNN-2D} & \textbf{ResNet} & \textbf{ResNet-2D} & \textbf{DAE} & \textbf{DAE-2D} & \textbf{E2CAR} & \textbf{E2CAR-2D} & \textbf{E2CAR-TPU} \\ 
\midrule
RRMSE-Time & 0.61 ± 0.08 & 0.78 ± 0.11 & 0.63 ± 0.17 & 0.61 ± 0.22 & 0.52 ± 0.10 & 0.52 ± 0.10 & 0.44 ± 0.10 & 0.54 ± 0.10 & 0.54 ± 0.10 \\
RRMSE-Freq & 0.61 ± 0.16 & 0.84 ± 0.08 & 0.43 ± 0.23 & 0.59 ± 0.23 & 0.53 ± 0.21 & 0.52 ± 0.21 & 0.37 ± 0.20 & 0.50 ± 0.20 & 0.50 ± 0.19 \\
CC & 0.80 ± 0.07 & 0.64 ± 0.15 & 0.79 ± 0.12 & 0.76 ± 0.19 & 0.86 ± 0.07 & 0.86 ± 0.07 & 0.90 ± 0.05 & 0.85 ± 0.07 & 0.84 ± 0.07 \\ 
\bottomrule
\end{tabular}
\end{table*}

\begin{table*}[ht]
\centering
\caption{Motion Artifact Removal - Comparison of RRMSE values and CC}
\label{table:motion_rrmse}

\begin{tabular}{@{}lccccccccccc@{}}
\toprule
\textbf{Metric} & \textbf{CNN} & \textbf{CNN-2D} & \textbf{ResNet} & \textbf{ResNet-2D} &
\textbf{DAE} & \textbf{DAE-2D} & \textbf{E2CAR} & \textbf{E2CAR-2D} & \textbf{E2CAR-TPU} \\ 
\midrule
RRMSE-Time & 0.73 $\pm$ 0.09 & 0.83 $\pm$ 0.12 & 0.64 $\pm$ 0.19 & 0.57 $\pm$ 0.21 &
0.71 $\pm$ 0.10 & 0.72 $\pm$ 0.10 & 0.61 $\pm$ 0.14 & 0.70 $\pm$ 0.11 & 0.70 $\pm$ 0.11 \\
RRMSE-Freq & 0.80 $\pm$ 0.14 & 0.86 $\pm$ 0.10 & 0.44 $\pm$ 0.19 & 0.56 $\pm$ 0.20 &
0.70 $\pm$ 0.17 & 0.69 $\pm$ 0.18 & 0.57 $\pm$ 0.17 & 0.72 $\pm$ 0.17 & 0.72 $\pm$ 0.17 \\
CC & 0.69 $\pm$ 0.10 & 0.54 $\pm$ 0.18 & 0.75 $\pm$ 0.20 & 0.78 $\pm$ 0.26 &
0.71 $\pm$ 0.11 & 0.70 $\pm$ 0.11 & 0.78 $\pm$ 0.14 & 0.71 $\pm$ 0.12 & 0.71 $\pm$ 0.12 \\ 
\bottomrule
\end{tabular}
\end{table*}

\begin{table*}[ht]
\centering
\label{table:emg_rrmse}
\caption{EMG Artifact Removal - Comparison of RRMSE values and CC}
\begin{tabular}{@{}lccccccccccc@{}}
\toprule
\textbf{Metric} & \textbf{CNN} & \textbf{CNN-2D} & \textbf{ResNet} & \textbf{ResNet-2D} & \textbf{DAE} & \textbf{DAE-2D} & \textbf{E2CAR} & \textbf{E2CAR-2D} & \textbf{E2CAR-TPU} \\ 
\midrule
RRMSE-Time & 0.66 ± 0.16 & 0.78 ± 0.14 & 0.63 ± 0.17 & 0.67 ± 0.24 & 0.59 ± 0.16 & 0.59 ± 0.17 & 0.53 ± 0.15 & 0.61 ± 0.17 & 0.61 ± 0.17 \\
RRMSE-Freq & 0.73 ± 0.17 & 0.88 ± 0.10 & 0.51 ± 0.20 & 0.69 ± 0.26 & 0.61 ± 0.18 & 0.60 ± 0.18 & 0.50 ± 0.18 & 0.65 ± 0.19 & 0.65 ± 0.19 \\
CC & 0.74 ± 0.14 & 0.62 ± 0.20 & 0.76 ± 0.16 & 0.69 ± 0.27 & 0.79 ± 0.12 & 0.79 ± 0.12 & 0.84 ± 0.10 & 0.79 ± 0.13 & 0.78 ± 0.13 \\ 
\bottomrule
\end{tabular}
\end{table*}

\textit{}

\section{Experiment}

\subsection{Model Implementation}
We adopted two basic deep learning model structures, ResNet\cite{he2016deep} and Autoencoder\cite{xing2024deep}, for code reproduction. We also reproduced another classic CNN model\cite{zhang2021eegdenoisenet} for comparison with our E2CAR model. The models are deployed to the Coral Dev board mini, thus the inference can be run on edge devices including CPU and edge TPU according to the method introduced previously. The model is converted to Tensorflow lite version. The test dataset, which is mentioned previously, is treated as the input of the model. The script loads the input testing dataset and transfers the data to the model. We tested several models on both the CPU and TPU. After running the inference script on Coral Dev board mini, the model's output results can be obtained and saved as a Numpy array. It can be later transferred to the laptop for further investigation. The data can be analyzed by calculating RRMSE and CC according to the method mentioned previously. Then we can understand the performance of the model more analytically, such as whether the output signal is closer to the original clean signal or not. 
\subsection{Coral Dev Board Mini Deployment}
In order to apply models to the Coral Dev board mini CPU, an inference script is necessary. We utilize TensorFlow runtime API, which can be used to run inference on the Coral Dev board mini.
\subsection{Results}
Using the evaluation metrics mentioned before, we can understand the performance of the model on four different tasks, including clean signal reconstruction, EOG artifact removal, EMG artifact removal and motion artifact removal.

\subsubsection{Clean EEG Signal Reconstruction}
In the task of clean EEG signal reconstruction, as shown in the first row in Table\ref{table:clean_rrmse}, the output of E2CAR-1D model has a smaller value in both time and frequency domain, whose mean values are about 0.24 and 0.2 respectively, indicating a more precious output compared to other models, including CNN, Res-Net and Autoencoder. Similarly to E2CAR-1D, E2CAR-TPU also has a small RRMSE in both domains. The cross correlation of E2CAR is higher than that of CNN, and a little bit larger than other three models.

\subsubsection{EOG Artifact Removal}
When it comes to EOG artifact removal, shown in the second row in Table\ref{table:eog_rrmse}, the E2CAR-1D model output also has the smallest RRMSE among all the models' output. And the  E2CAR-TPU model output has a similar performance compared to its 1D version. It is worth noting that the CC value of E2CAR model is significantly higher than those of other models, indicating a great improvement in model output, and it is highly correlated with the true clean signal. 
\subsubsection{EMG Artifact Removal}
The evaluation results of models related to EMG artifact removal are shown in the third row in Table \ref{table:emg_rrmse}.  E2CAR still has the best performance among all the models, including both RRMSE and CC. It is worth noting that the performance of E2CAR is similar when using CPU and TPU, which means the model structure is better supported in edge devices like Coral Dev board mini.
\subsubsection{Motion Artifact Removal}
The last row of Table \ref{table:motion_rrmse} shows the evaluation result of motion artifact removal. The result is still good, although RRMSE in frequency domain of E2CAR is a little bit higher than that of Autoencoder structure, which means the model may capture some useless feature of motion artifact, thus leading to a bad performance on this task.

\subsubsection{Coral Dev Board Mini Computational Cost}
The computational cost on edge devices is the most important aspect of evaluating our method since we should deploy an easy model, which can infer fast on edge devices while having good results. As shown in the first column of table\ref{table:computational cost}, when expanding model dimension from 1D to 2D, the inference time of the model will drop from about 38 percent to 50 percent. When further applied to the edge TPU after compiling, inference speed of all the models improves, and E2CAR uses about 5.7ms to process a 4s-input-signal. Which is enough for real-time tasks.
The second column of table \ref{table:computational cost} shows the power consumption of different models. All the models have less energy consumption after optimizing. Also, it is worth noting that although E2CAR has the most complex model structure, it has the least power consumption compared to other simple deep learning models. What's more, the power consumption of E2CAR on TPU is the lowest, about 0.36mAh/h.

\begin{table}[ht]
\centering
\caption{Power Consumption and Inference Time Comparison}
\label{table:computational cost}
\begin{tabular}{@{}lcc@{}}
\toprule
\textbf{Model Type} & \textbf{Inference Time (ms)} & \textbf{Power Consumption (mAh/h)} \\ \midrule
\textbf{CNN}        &                              &                                  \\
1D                  & 37.4                         & 0.48                             \\
2D                  & 19.7                         & 0.41                             \\
TPU                 & 1.7                          & 0.39                             \\ \midrule
\textbf{ResNet}     &                              &                                  \\
1D                  & 141.7                        & 0.45                             \\
2D                  & 69.8                         & 0.39                             \\
TPU                 & 5.3                          & 0.30                             \\ \midrule
\textbf{DAE}        &                              &                                  \\
1D                  & 21.2                         & 0.43                             \\
2D                  & 13.4                         & 0.39                             \\
TPU                 & 2.1                          & 0.34                             \\ \midrule
\textbf{E2CAR}      &                              &                                  \\
1D                  & 146.0                        & 0.41                             \\
2D                  & 91.1                         & 0.38                             \\
TPU                 & 5.7                          & 0.36                             \\ \bottomrule
\end{tabular}
\end{table}

\section{Discussions}
\subsection{Generalizability and Robustness}
E2CAR is a simple CNN network, which can be deployed on resource-limited edge devices for real-time application. Unlike other simple CNN models which can only remove specific types of artifacts, the encoder of the model can capture various artifacts, making it an efficient one-in-all model, which is better than using different model for different tasks. This means it can complete different tasks using one model, reducing the consumption of the memory usage of edge devices, which can be actually be used in real-life tasks.

The method E2CAR used, expanding 1-D convolutional network to 2-D, can be applied in most of the CNN models using convolution operation to reduce inference time and power consumption. We modified CNN, Res-net, and Autoencoder models, the result of which is shown in Table \ref{table:computational cost}. This method can be applied generally on different models to achieve a better inference speed when applied to edge devices like edge TPU.
 
\subsection{Future work}
Since the model has successfully applied to the edge device, it can be used in real-world applications in the future. Coral Dev board mini can be connected to the sensors to collect EEG signals and remove artifacts directly. And the outcome is transmitted through a wifi module to the laptop or servers to provide a real-time, clean EEG signal, making it a complete EEG artifact removal IoT system.

\subsection{Conclusion}
In this study, we proposed the E2CAR framework, a 2D convolutional neural network optimized for real-time EEG artifact removal, designed for edge devices. We introduced a residual module to address the vanishing gradient problem in DAE. By expanding the model from 1D to 2D convolution and deploying it on the Coral Dev Board mini’s TPU, we significantly improved inference time and reduced power consumption, while maintaining high artifact removal accuracy. Experimental results show that E2CAR outperforms traditional simple model structures, making it an ideal solution for real-time EEG processing on resource-constrained devices. Future work will focus on integrating the framework into a complete IoT system for practical applications.

\bibliographystyle{IEEEtran}
\bibliography{main}

@inproceedings{li2025human,
title={{Human Motion Instruction Tuning}},
author={Li, Lei and Jia, Sen and Wang, Jianhao and Jiang, Zhongyu and Zhou, Feng and Dai, Ju and Zhang, Tianfang and Wu, Zongkai and Hwang, Jenq-Neng},
booktitle={Proceedings of the IEEE/CVF Conference on Computer Vision and Pattern Recognition (CVPR)},
year={2025}
}

@inproceedings{yao2025countllm,
title={{CountLLM: Towards Generalizable Repetitive Action Counting via Large Language Model}},
author={Yao, Ziyu and Cheng, Xuxin and Huang, Zhiqi and Li, Lei},
booktitle={Proceedings of the IEEE/CVF Conference on Computer Vision and Pattern Recognition (CVPR)},
year={2025}
}

@inproceedings{cai2025bayesian,
title={{Bayesian Optimization for Controlled Image Editing via LLMs}},
author={Cai, Chengkun and Liu, Haoliang and Zhao, Xu and Jiang, Zhongyu and Zhang, Tianfang and Wu, Zongkai and Lee, John and Hwang, Jenq-Neng and Li, Lei},
booktitle={Proceedings of the Annual Meeting of the Association for Computational Linguistics (ACL)},
year={2025}
}

@inproceedings{cai2025role,
title={{The Role of Deductive and Inductive Reasoning in Large Language Models}},
author={Cai, Chengkun and Zhao, Xu and Liu, Haoliang and Jiang, Zhongyu and Zhang, Tianfang and Wu, Zongkai and Hwang, Jenq-Neng and Li, Lei},
booktitle={Proceedings of the Annual Meeting of the Association for Computational Linguistics (ACL)},
year={2025}
}

@inproceedings{shi2024scaling,
title={{Scaling Law for Time Series Forecasting}},
author={Shi, Jingzhe and Ma, Qinwei and Ma, Huan and Li, Lei},
booktitle={Advances in Neural Information Processing Systems (NeurIPS)},
year={2024}
}

@inproceedings{gu2025mocount,
title={MoCount: Motion-Based Repetitive Action Counting},
author={Gu, Ruocheng and Jia, Sen and Ma, Yule and Zhong, Jinqin and Hwang, Jenq-Neng and Li, Lei},
booktitle={Proceedings of the 33rd ACM International Conference on Multimedia},
pages={9026--9034},
year={2025}
}

@inproceedings{lan-etal-2025-attention,
title = "Attention Consistency for {LLM}s Explanation",
author = "Lan, Tian  and
Xu, Jinyuan  and
He, Xue  and
Hwang, Jenq-Neng  and
Li, Lei",
editor = "Christodoulopoulos, Christos  and
Chakraborty, Tanmoy  and
Rose, Carolyn  and
Peng, Violet",
booktitle = "Findings of the Association for Computational Linguistics: EMNLP 2025",
month = nov,
year = "2025",
address = "Suzhou, China",
publisher = "Association for Computational Linguistics",
url = "https://aclanthology.org/2025.findings-emnlp.91/",
doi = "10.18653/v1/2025.findings-emnlp.91",
pages = "1736--1750",
ISBN = "979-8-89176-335-7"
}

@article{chaddad2023electroencephalography,
  title={Electroencephalography signal processing: A comprehensive review and analysis of methods and techniques},
  author={Chaddad, Ahmad and Wu, Yihang and Kateb, Reem and Bouridane, Ahmed},
  journal={Sensors},
  volume={23},
  number={14},
  pages={6434},
  year={2023},
  publisher={MDPI}
}

@article{jiang2019removal,
  title={Removal of artifacts from EEG signals: a review},
  author={Jiang, Xiao and Bian, Gui-Bin and Tian, Zean},
  journal={Sensors},
  volume={19},
  number={5},
  pages={987},
  year={2019},
  publisher={MDPI}
}

@inproceedings{djuwari2006limitations,
  title={Limitations of ICA for artefact removal},
  author={Djuwari, Djuwari and Kumar, Dinesh Kant and Palaniswami, Marimuthu},
  booktitle={2005 IEEE Engineering in Medicine and Biology 27th Annual Conference},
  pages={4685--4688},
  year={2006},
  organization={IEEE}
}

@inproceedings{naik2006limitations,
  title={Limitations and applications of ICA for surface electromyogram},
  author={Naik, Ganesh R and Kumar, Dinesh K and Arjunan, Sridhar P and Palaniswami, Marimuthu and Begg, Rezaul},
  booktitle={2006 International Conference of the IEEE Engineering in Medicine and Biology Society},
  pages={5739--5742},
  year={2006},
  organization={IEEE}
}

@article{mumtaz2021review,
  title={Review of challenges associated with the EEG artifact removal methods},
  author={Mumtaz, Wajid and Rasheed, Suleman and Irfan, Alina},
  journal={Biomedical Signal Processing and Control},
  volume={68},
  pages={102741},
  year={2021},
  publisher={Elsevier}
}

@article{yang2018automatic,
  title={Automatic ocular artifacts removal in EEG using deep learning},
  author={Yang, Banghua and Duan, Kaiwen and Fan, Chengcheng and Hu, Chenxiao and Wang, Jinlong},
  journal={Biomedical Signal Processing and Control},
  volume={43},
  pages={148--158},
  year={2018},
  publisher={Elsevier}
}

@article{stalin2021machine,
  title={A Machine Learning-Based Big EEG Data Artifact Detection and Wavelet-Based Removal: An Empirical Approach},
  author={Stalin, Shalini and Roy, Vandana and Shukla, Prashant Kumar and Zaguia, Atef and Khan, Mohammad Monirujjaman and Shukla, Piyush Kumar and Jain, Anurag},
  journal={Mathematical Problems in Engineering},
  volume={2021},
  number={1},
  pages={2942808},
  year={2021},
  publisher={Wiley Online Library}
}

@inproceedings{mashhadi2020deep,
  title={Deep learning denoising for EOG artifacts removal from EEG signals},
  author={Mashhadi, Najmeh and Khuzani, Abolfazl Zargari and Heidari, Morteza and Khaledyan, Donya},
  booktitle={2020 IEEE Global Humanitarian Technology Conference (GHTC)},
  pages={1--6},
  year={2020},
  organization={IEEE}
}

@article{zhang2020deep,
  title={Deep learning in the era of edge computing: Challenges and opportunities},
  author={Zhang, Mi and Zhang, Faen and Lane, Nicholas D and Shu, Yuanchao and Zeng, Xiao and Fang, Biyi and Yan, Shen and Xu, Hui},
  journal={Fog Computing: Theory and Practice},
  pages={67--78},
  year={2020},
  publisher={Wiley Online Library}
}

@article{chen2020deep,
  title={Deep learning on mobile and embedded devices: State-of-the-art, challenges, and future directions},
  author={Chen, Yanjiao and Zheng, Baolin and Zhang, Zihan and Wang, Qian and Shen, Chao and Zhang, Qian},
  journal={ACM Computing Surveys (CSUR)},
  volume={53},
  number={4},
  pages={1--37},
  year={2020},
  publisher={ACM New York, NY, USA}
}

@article{klados2016semi,
  title={A semi-simulated EEG/EOG dataset for the comparison of EOG artifact rejection techniques},
  author={Klados, Manousos A and Bamidis, Panagiotis D},
  journal={Data in brief},
  volume={8},
  pages={1004--1006},
  year={2016},
  publisher={Elsevier}
}

@article{sweeney2012methodology,
  title={A methodology for validating artifact removal techniques for physiological signals},
  author={Sweeney, Kevin T and Ayaz, Hasan and Ward, Tom{\'a}s E and Izzetoglu, Meltem and McLoone, Se{\'a}n F and Onaral, Banu},
  journal={IEEE transactions on information technology in biomedicine},
  volume={16},
  number={5},
  pages={918--926},
  year={2012},
  publisher={IEEE}
}

@article{zhang2021eegdenoisenet,
  title={EEGdenoiseNet: a benchmark dataset for deep learning solutions of EEG denoising},
  author={Zhang, Haoming and Zhao, Mingqi and Wei, Chen and Mantini, Dante and Li, Zherui and Liu, Quanying},
  journal={Journal of Neural Engineering},
  volume={18},
  number={5},
  pages={056057},
  year={2021},
  publisher={IOP Publishing}
}

@inproceedings{he2016deep,
  title={Deep residual learning for image recognition},
  author={He, Kaiming and Zhang, Xiangyu and Ren, Shaoqing and Sun, Jian},
  booktitle={Proceedings of the IEEE conference on computer vision and pattern recognition},
  pages={770--778},
  year={2016}
}

@article{xing2024deep,
  title={Deep autoencoder for real-time single-channel EEG cleaning and its smartphone implementation using TensorFlow Lite with hardware/software acceleration},
  author={Xing, Le and Casson, Alexander J},
  journal={IEEE Transactions on Biomedical Engineering},
  year={2024},
  publisher={IEEE}
}

@article{vigario1997extraction,
  title={Extraction of ocular artefacts from EEG using independent component analysis},
  author={Vig{\'a}rio, Ricardo Nuno},
  journal={Electroencephalography and clinical neurophysiology},
  volume={103},
  number={3},
  pages={395--404},
  year={1997},
  publisher={Elsevier}
}

@article{zhuang2020technical,
  title={A technical review of canonical correlation analysis for neuroscience applications},
  author={Zhuang, Xiaowei and Yang, Zhengshi and Cordes, Dietmar},
  journal={Human brain mapping},
  volume={41},
  number={13},
  pages={3807--3833},
  year={2020},
  publisher={Wiley Online Library}
}

@article{fitzgibbon2007removal,
  title={Removal of EEG noise and artifact using blind source separation},
  author={Fitzgibbon, S P and Powers, D MW and Pope, K J and Clark, C R},
  journal={Journal of Clinical Neurophysiology},
  volume={24},
  number={3},
  pages={232--243},
  year={2007},
  publisher={LWW}
}

@article{yu2022edge,
  title={Edge computing-assisted IoT framework with an autoencoder for fault detection in manufacturing predictive maintenance},
  author={Yu, Wenjin and Liu, Yuehua and Dillon, Tharam and Rahayu, Wenny},
  journal={IEEE Transactions on Industrial Informatics},
  volume={19},
  number={4},
  pages={5701--5710},
  year={2022},
  publisher={IEEE}
}

@inproceedings{canziani2017evaluation,
  title={Evaluation of neural network architectures for embedded systems},
  author={Canziani, Alfredo and Culurciello, Eugenio and Paszke, Adam},
  booktitle={2017 IEEE international symposium on Circuits and systems (ISCAS)},
  pages={1--4},
  year={2017},
  organization={IEEE}
}

@article{garcia2023analysing,
  title={Analysing edge computing devices for the deployment of embedded AI},
  author={Garcia-Perez, Asier and Mi{\~n}{\'o}n, Ra{\'u}l and Torre-Bastida, Ana I and Zulueta-Guerrero, Ekaitz},
  journal={Sensors},
  volume={23},
  number={23},
  pages={9495},
  year={2023},
  publisher={MDPI}
}

@inproceedings{baller2021deepedgebench,
  title={DeepEdgeBench: Benchmarking deep neural networks on edge devices},
  author={Baller, Stephan Patrick and Jindal, Anshul and Chadha, Mohak and Gerndt, Michael},
  booktitle={2021 IEEE International Conference on Cloud Engineering (IC2E)},
  pages={20--30},
  year={2021},
  organization={IEEE}
}

\end{document}